\documentclass[aps,prb,reprint,showpacs,amsfonts,amssymb,amsmath,floatfix]{revtex4-1}
\usepackage[pdftex]{graphicx}
\usepackage[bookmarks=false,colorlinks=true,linkcolor=blue,citecolor=blue,urlcolor=blue,breaklinks]{hyperref}

\newcommand{\Tr}{\operatorname{Tr}}

\def\wt#1{\widetilde{#1}}

\def\ua{\uparrow}
\def\da{\downarrow}

\def\one{\hat{1}}

\def\Br{\mathcal{B}}
\def\Hm{\mathcal{H}}

\def\ve{\varepsilon}
\def\vp{\varphi}
\def\vt{\vartheta}

\def\Sam{\mathrm{S}}

\def\sigv{\boldsymbol{\sigma}}
\def\Omegv{\boldsymbol{\Omega}}

\def\Bv{\mathbf{B}}

\def\Gh{\hat{G}}
\def\Ih{\hat{I}}

\def\kv{\mathbf{k}}

\def\qv{\mathbf{q}}
\def\rv{\mathbf{r}}
\def\Rv{\mathbf{R}}
\def\rhoh{\hat{\rho}}

\def\sigh{\hat{\sigma}}
\def\Sigh{\hat{\Sigma}}
\def\sv{\mathbf{s}}
\def\Sv{\mathbf{S}}

\def\zh{\hat{\mathbf{z}}}

\begin{document}
\title{Spin relaxation near a ferromagnetic transition}
\author{Matthew D. Mower}
\affiliation{Department of Physics, Truman State University, Kirksville, Missouri 63501, USA}
\author{G. Vignale}
\affiliation{Department of Physics and Astronomy, University of Missouri, Columbia, Missouri 65211, USA}

\begin{abstract}
We study spin relaxation in dilute magnetic semiconductors near a ferromagnetic transition, where spin fluctuations become strong. An enhancement in the scattering rate of itinerant carriers from the spin fluctuations of localized impurities leads to a change in the dominant spin relaxation mechanism from Dyakonov-Perel to spin flips in scattering. On the ferromagnetic side of the transition, we show that due to the presence of two magnetic components -- the itinerant carriers and the magnetic impurities -- with different gyromagnetic ratios, the relaxation rate of the total magnetization can be quite different from the relaxation rate of the spin. Following a disturbance of the equilibrium magnetization, the spin is initially \textit{redistributed} between the two components to restore the equilibrium magnetization. It is only on a longer time scale, controlled by the spin-orbit interaction, that the total spin itself relaxes to its equilibrium state.
\end{abstract}

\pacs{}

\maketitle

\section{Introduction}
The draw of magnetic semiconductors is readily apparent from the name; they offer an opportunity to blend magnetic and semiconducting properties in a single material. In spintronics, this is especially relevant because of the additional means of interaction through the magnetic and electrical properties of carriers. Dilute magnetic semiconductors\cite{RevModPhys.78.809,nmat1325,PhysRevLett.68.2664,PhysRevLett.63.1849,ohno:363} are unique in that they marry a low-magnetism ferromagnet with a semiconductor capable of maintaining long spin lifetimes. This makes possible the manipulation of carrier spins by ferromagnetic switching or, in the non-magnetic state, through more traditional means like optical excitation. While these materials are a reality, the ferromagnetic transitions are so far confined to well below room temperature.\cite{nmat2913,nmat2898,RevModPhys.86.855} Nonetheless, the potential for these materials warrants their study, not to mention their suitability for studying spin systems near critical transitions, like a ferromagnetic transition.

In this paper, we examine the impact of a ferromagnetic transition on carrier spin lifetimes. Strong spin fluctuations are expected for the constituents driving a ferromagnetic phase change.\cite{RevModPhys.79.1015} These fluctuations can manifest themselves through enhanced carrier scattering. In the Dyakonov-Perel spin relaxation mechanism,\cite{dyak_1} spin lifetimes are typically inversely related to scattering lifetimes. This is because the relaxation process occurs between collisions, due to precession of the spin in the spin-orbit field. Then, in the Dyakonov-Perel mechanism, enhanced scattering due to spin fluctuations disrupts spin precession and can inhibit the rate of spin relaxation. Under these circumstances, we can find a crossover in the dominant spin relaxation mechanism to one based on spin flips in scattering. Spin flips can occur in scattering, for example, due to the mixing of spin-up and spin-down states from a spin-orbit interaction; this is the premise of the Elliott-Yafet spin relaxation mechanism.\cite{PhysRev.85.478,*PhysRev.96.266,Wu201061} In a ferromagnetic system, an even greater rate of spin flips in interaction is made possible by transfer of spin between itinerant carriers and magnetic impurities.

We model a ferromagnetic transition in a dilute magnetic semiconductor with a degenerate gas of itinerant carriers (electrons or holes) interacting via spin exchange with dilute localized magnetic impurities. {GaMnAs} is the prototypical material to which this model has been applied -- not without controversy over the hole transport mechanism.\cite{nmat3250,PhysRevLett.105.227202,nmat2898} {GaGdN} is another example: here it is the $s$-$f$ exchange interaction (as opposed to $s$-$d$ interaction) that drives the ferromagnetic transition.\cite{PhysRevB.72.115201,PSSC:PSSC200778560,Wu201061} In this paper, we will keep the discussion fairly general, applying parameters related to a {GaMnAs} system for the sake of calculations and analysis, but noting that the model is meant more as an example of critical phenomena than an exact description of {GaMnAs}. Other articles have examined the critical behavior of {GaMnAs} in more detail.\cite{PhysRevLett.101.077201,PhysRevB.72.035210,PhysRevLett.105.227202,PhysRevB.80.205202,PhysRevB.83.205206,Dietl11022000} Our treatment is mean-field-like in that spin fluctuations and the interaction they mediate between the carriers are at the level of a random phase approximation (RPA). Thus, we ignore critical fluctuations which presumably lead to large corrections to the quasiparticle spectrum in the immediate vicinity of the ferromagnetic transition.

Derivations of spin relaxation times resulting from spin dependent scattering are presented for two mechanisms: transfer of spin from carriers to magnetic impurities and Dyakonov-Perel. The results we obtain are applicable to dynamic, spin-polarized systems and focus primarily on the contribution of carrier-carrier interactions mediated by the spin fluctuations of magnetic impurities. We apply common Fermi liquid theory techniques, beginning with the Kadanoff-Baym kinetic equation. By utilizing the GW approximation for the carrier self-energy, we obtain relatively simple and easy to calculate expressions for the spin relaxation rate, which are generally applicable to a wide variety of effective interactions. The results derived here follow previous analytic\cite{Wu201061,jetp_99_6_1279,PhysRevB.68.075312,jetpl_75_8_403,PhysRevB.83.155205} and computational\cite{PhysRevB.79.155201,PhysRevB.85.075206,PhysRevB.79.125206} studies, but extend them significantly by including dynamics and spin exchange.

This paper is organized as follows: in Section~\ref{EFF_INT}, we present an effective interaction that can be used to generate a ferromagnetic transition and which will be used to calculate carrier scattering rates; in Section~\ref{FMP}, we discuss the ferromagnetic phase change and the behavior of the scattering amplitude near the transition; in Section~\ref{SR} we derive analytic expressions for the rate of spin relaxation due to spin-flips in scattering and the Dyakonov-Perel mechanism; in Section~\ref{DISCUSSION}, we discuss some results of spin relaxation across a ferromagnetic transition; Section~\ref{CONCLUSION} contains our concluding remarks.

\section{Effective spin-spin interaction}
\label{EFF_INT}
We first introduce the model dilute magnetic semiconductor on which we base calculations. Our model resembles the Zener model of ferromagnetism, with a ferromagnetic transition being driven by spin-exchange interactions between itinerant carriers and localized magnetic impurities.\cite{Dietl11022000,PhysRev.81.440} The Hamiltonian includes contributions from carriers $\Hm_{\sigma}$, magnetic impurities $\Hm_S$, and interactions between carriers and magnetic impurities $\Hm_{\sigma S}$:
\begin{subequations}
\begin{align}
	\label{hamiltonian_carriers}
	\Hm_{\sigma} &=\sum_i \left\{\frac{p_i^2}{2m^*}+ \mu_B g_{\sigma} \sigv_i\cdot \Bv_{\sigma}(\rv_i)\right\}, \\
	\label{hamiltonian_impurities}
	\Hm_S &= \mu_Bg_S \sum_j \Sv_j \cdot \Bv_{S}(\Rv_j), \\
	\label{hamiltonian_interaction}
	\Hm_{\sigma S} &= J\sum_{i,j}\sigv_i\cdot \Sv_j \delta(\rv_i-\Rv_j),
\end{align}
\end{subequations}
where $\rv_i$ and $\sigv_i$ are, respectively, the positions and spins of the mobile carriers with effective mass $m^*$, $\Rv_j$ and $\Sv_j$ are the positions and spins of the fixed impurities, $\mu_B$ is the Bohr magneton, $g_{\sigma}$ and $g_S$ are the $g$-factors which determine the magnetic moments of the carriers and the magnetic impurities as $\mu_B g_{\sigma}$ and $\mu_B g_S$, respectively, and $\Bv_{\sigma}$ and $\Bv_S$ are magnetic fields which are supposed to act only on carriers or magnetic impurities, respectively. The contact-type interaction between carriers and magnetic impurities is represented by $J\delta(\rv_i-\Rv_j)$, where $J$ is the interaction strength averaged over the unit cell. The impurities are taken to be sufficiently dilute that they do not interact with each other directly.

The alignment of spins with an external magnetic field $\Bv$ is linearized as $\mu_Bg\langle \sv \rangle = \chi \langle\Bv\rangle$, where $\chi$ is the magnetic susceptibility. By combining the interaction $J$ with the magnetic fields, we can derive effective mean fields $\Bv^{\text{eff}}_{\sigma}$ and $\Bv^{\text{eff}}_S$ acting on carriers and impurities, respectively:
\begin{align}
	\label{Beff_sigma}
	\mu_Bg_{\sigma} \Bv^{\text{eff}}_{\sigma}(\rv) &= \mu_Bg_{\sigma} \Bv_{\sigma}(\rv) + J \Sv(\rv), \\
	\label{Beff_S}
	\mu_Bg_S \Bv^{\text{eff}}_S(\Rv) &= \mu_Bg_S \Bv_{S}(\Rv) + J \sigv(\Rv),
\end{align}
where $\sigv(\rv) \equiv \sum_i \delta(\rv-\rv_i) \sigv_i$ and $\Sv(\rv) \equiv \sum_j \delta(\Rv-\Rv_j) \Sv_j$ are the densities of carrier spin evaluated at $\rv$ and impurity spin evaluated at $\Rv$, respectively. We are not including spin-orbit in the effective carrier field due to its relatively small strength compared to $J$. The magnetic fields in Eqs.~\eqref{Beff_sigma} and \eqref{Beff_S} are specific to each species, so it is cleanest to absorb $\mu_Bg$ into these fields and write
\begin{align}
	\label{susc_car}
	\langle\sigv\rangle &= \chi^{(0)}_{\sigma\sigma} \langle\Bv^{\text{eff}}_{\sigma}\rangle, \\
	\label{susc_imp}
	\langle\Sv\rangle &= \chi^{(0)}_{SS} \langle\Bv^{\text{eff}}_S\rangle,
\end{align}
where $\chi^{(0)}_{\sigma\sigma}$ and $\chi^{(0)}_{SS}$ are non-interacting (with respect to $J$) spin-spin susceptibilities for carriers and magnetic impurities, respectively. They have dimensions of inverse energy-volume. To be explicit, these fields relate to the physical magnetic field by $\Bv_s=g_s\mu_B \Bv$ and the susceptibilities are related to textbook magnetic susceptibilities by $\chi_{ss}=\chi/(g\mu_B)^2$.

The equilibrium spin polarization of the impurities $\langle S_z\rangle$ follows the Brillouin function $\Br_J(x)$,\cite{kittel_issp}
\begin{equation}
	\label{Sz_brillouin}
	\langle S_z \rangle = -\Sam_S n_S \Br_{\Sam_S}\left[\frac{\Sam_S \langle B_{S,z} \rangle}{k_BT}\right],
\end{equation}
where $\Sam_S$ is the magnitude of the impurity spin, $n_S$ is the density of magnetic impurities, and $T$ is the temperature. Then, the longitudinal and transverse components of the spin susceptibility of the impurities are calculated from
\begin{align}
	\chi^{(0)}_{S_zS_z} &= \partial \langle S_z\rangle/\partial \langle B_{S,z} \rangle, \\
	\chi^{(0)}_{S_{\pm}S_{\mp}} &= 2\langle S_z \rangle/\langle B_{S,z} \rangle.
\end{align}
The non-interacting spin response of carriers coincides with Lindhard response:
\begin{align}
	\chi^{(0)}_{\sigma_z\sigma_z}(q,\omega) &= \Sam_{\sigma}^2 \sum_{\kv, \alpha} \frac{f^0_{k \alpha} - f^0_{\kv+\qv \alpha}}{\hbar\omega + \ve_{k\alpha} - \ve_{\kv+\qv \alpha}+i \eta}, \\
	\chi^{(0)}_{\sigma_+\sigma_-}(q,\omega) &= 4 \Sam_{\sigma}^2 \sum_{\kv} \frac{f^0_{k \ua} - f^0_{\kv+\qv \da}}{\hbar\omega + \ve_{k\ua} - \ve_{\kv+\qv \da}+i \eta},
\end{align}
where $\Sam_{\sigma}$ is the magnitude of the carrier spin, $f^0_{k\alpha}$ is the equilibrium Fermi-Dirac distribution, and $\chi^{(0)}_{\sigma_-\sigma_+}(q,\omega) = [\chi^{(0)}_{\sigma_+\sigma_-}(q,-\omega)]^*$.

We can also shift the interaction $J$ into the spin susceptibilities to generate effective response functions. This is accomplished by solving the coupled linear Eqs.~\eqref{susc_car} and \eqref{susc_imp} for $\langle \sigv \rangle$ and $\langle \Sv \rangle$ and identifying the effective response function as the proportionality between average spin and bare field:
\begin{equation}
\begin{bmatrix}
	\langle S_z \rangle \\
	\langle \sigma_z \rangle	
\end{bmatrix} = \underline{\underline{\chi_l}}
\begin{bmatrix}
	\langle B_{S,z} \rangle \\
	\langle B_{\sigma,z} \rangle
\end{bmatrix},
\end{equation}
and
\begin{equation}
\begin{bmatrix}
	\langle S_+ \rangle \\
	\langle S_- \rangle \\
	\langle \sigma_+ \rangle \\
	\langle \sigma_- \rangle
\end{bmatrix} = 
\frac{1}{2}\underline{\underline{\chi_t}}
\begin{bmatrix}
	\langle B_{S+} \rangle \\
	\langle B_{S-} \rangle \\
	\langle B_{\sigma+} \rangle \\
	\langle B_{\sigma-} \rangle
\end{bmatrix},
\end{equation}
where $s_{\pm} = s_x \pm i s_y$ and $B_{s\pm} = B_{s,x} \pm i B_{s,y}$. The effective susceptibilities are found to be
\begin{equation}
	\label{chi_long}
	\underline{\underline{\chi_l}} = 
\begin{bmatrix}
	\chi_{S_zS_z} & \chi_{S_z\sigma_z} \\
	\chi_{\sigma_z S_z} & \chi_{\sigma_z\sigma_z}
\end{bmatrix} =
\begin{bmatrix}
	1/\chi_{S_zS_z}^{(0)} & -J \\
	-J & 1/\chi_{\sigma_z\sigma_z}^{(0)}
\end{bmatrix}^{-1}
\end{equation}
and
\begin{align}
\label{chi_trans}
\underline{\underline{\chi_t}} &= \begin{bmatrix}
	\chi_{S_+S_-} & \chi_{S_+S_+} & \chi_{S_+\sigma_-} & \chi_{S_+\sigma_+} \\
	\chi_{S_-S_-} & \chi_{S_-S_+} & \chi_{S_-\sigma_-} & \chi_{S_-\sigma_+} \\
	\chi_{\sigma_+S_-} & \chi_{\sigma_+S_+} & \chi_{\sigma_+\sigma_-} & \chi_{\sigma_+\sigma_+} \\
	\chi_{\sigma_-S_-} & \chi_{\sigma_-S_+} & \chi_{\sigma_-\sigma_-} & \chi_{\sigma_-\sigma_+}
\end{bmatrix} \nonumber \\
	&= \begin{bmatrix}
	1/\chi_{S_+S_-}^{(0)} & 0 & -J/2 & 0 \\
	0 & 1/\chi_{S_-S_+}^{(0)} & 0 & -J/2 \\
	-J/2 & 0 & 1/\chi_{\sigma_+\sigma_-}^{(0)} & 0 \\
	0 & -J/2 & 0 & 1/\chi_{\sigma_-\sigma_+}^{(0)}
\end{bmatrix}^{-1}.
\end{align}
Rather than give explicit expressions for every effective response function in Eqs.~\eqref{chi_long} and \eqref{chi_trans}, we write only those which are used in this paper. Namely, the effective impurity spin-spin response functions are
\begin{align}
	\label{chiSzSzeff}
	\chi_{S_zS_z}(q,\omega) &= \frac{\chi^{(0)}_{S_zS_z}}{1-J^2 \chi^{(0)}_{S_zS_z} \chi^{(0)}_{\sigma_z\sigma_z}(q,\omega)}, \\
	\label{chiSpSmeff}
	\chi_{S_+S_-}(q,\omega) &= \frac{\chi^{(0)}_{S_+S_-}}{1-(J/2)^2 \chi^{(0)}_{S_+S_-}\chi^{(0)}_{\sigma_+\sigma_-}(q,\omega)},
\end{align}
where $\chi_{S_-S_+}(q,\omega)=[\chi_{S_+S_-}(q,-\omega)]^*$.

\begin{figure}
\begin{center}
	\includegraphics[width=4cm]{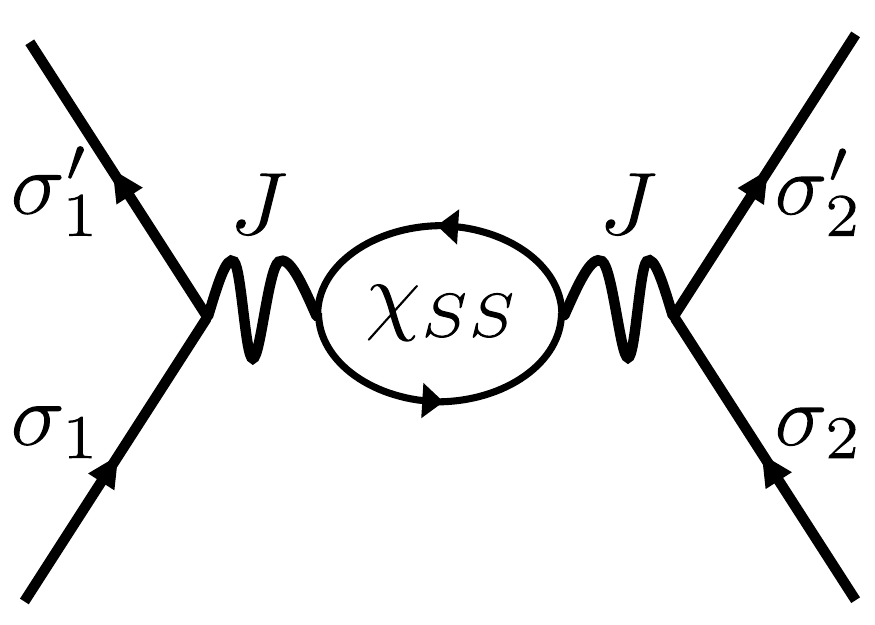}
\end{center}
\caption[Effective spin-spin interaction between carriers]{The effective spin-spin interaction between carriers. While carriers do not interact directly with each other via spin, the spin interaction $J$ can be mediated by impurities through the effective spin-spin susceptibility $\chi_{SS}$. Here, $\sigma$ and $\sigma'$ are the incoming and outgoing carrier spins, respectively.}
\label{fig:FeynCC}
\end{figure}

The interaction of interest to us is the effective carrier-carrier interaction, shown schematically in Fig.~\ref{fig:FeynCC}. By using the effective spin-spin susceptibility for impurities in Eqs.~\eqref{chiSzSzeff}-\eqref{chiSpSmeff}, we capture the effect of \textit{collective} impurity spin fluctuations on carrier-carrier scattering. Notice that although the non-interacting impurity spin response is frequency independent, the effective spin response is dynamic due to the inclusion of carrier dynamics. The longitudinal ($\parallel$) and transverse ($\perp$) components of this interaction are
\begin{align}
	\label{Wpar}
	W_{\parallel\qv}(\omega) &= J^2\chi_{S_zS_z}(q,\omega), \\
	\label{Wperp}
	W_{\perp\qv}(\omega) &= (J/2)^2\left[\chi_{S_+S_-}(q,\omega)+\chi_{S_-S_+}(q,\omega)\right].
\end{align}
It can be verified that these effective interactions are equivalent to an infinite sum of all possible bare impurity-carrier interactions mediated by non-interacting susceptibilities, e.g.
\begin{equation*}
	W = \sum_{n=1}^{\infty} J^{2n} \left[\chi^{(0)}_{SS}\right]^n \left[\chi^{(0)}_{\sigma\sigma}\right]^{n-1}.
\end{equation*}

\section{Ferromagnetic transition and phase}
\label{FMP}
The interaction $J$ between carriers and impurities allows for the possibility of non-zero average spins $\langle \sigv \rangle$ and $\langle \Sv \rangle$ in the absence of external fields. This is confirmed by the Curie-Weiss-type susceptibility to which the effective response functions in Eqs.~\eqref{chiSzSzeff} and \eqref{chiSpSmeff} reduce in the static, long wavelength limit: $\chi \sim (T-T_c)^{-1}$. The paramagnetic-ferromagnetic phase change occurs when
\begin{align}
	\label{Tc}
	&1 - J^2 \chi^{(0)}_{S_zS_z}(T_c)\lim_{q\rightarrow 0}\chi^{(0)}_{\sigma_z\sigma_z}(q,0) = 0 \nonumber \\
	&\quad \Rightarrow \quad T_c = \frac{\Sam_{S}(\Sam_{S}+1)}{3k_B} J^2 n_S \Sam_{\sigma}^2 N(0),
\end{align}
where $N(0)$ is the carrier density of states at the Fermi level.

\begin{figure}[t]
\begin{center}
	\includegraphics[width=0.48\textwidth]{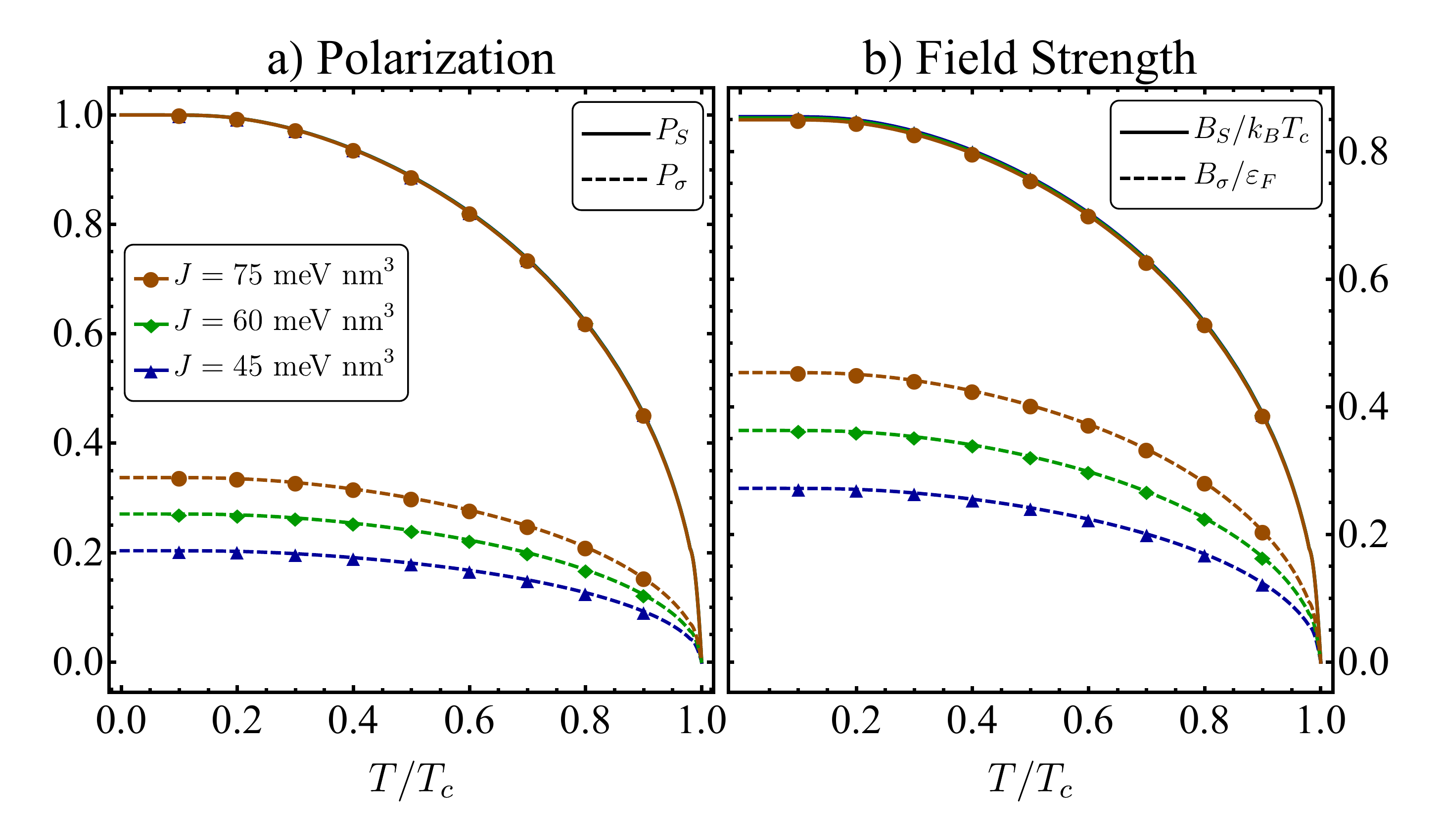}
\end{center}
\caption[Polarizations and fields]{(Color online) Mean polarizations of, and effective magnetic fields acting on carriers and magnetic impurities have been calculated from simultaneous solution of Eqs.~\eqref{polS}-\eqref{goldstone}. {GaMnAs} parameters have been used to define the carriers as holes with effective mass $m^* = 0.5 m_e$ and density $n_{\sigma} = 5\times 10^{20} \text{ cm}^{-3}$. The Mn acceptors have spin $\Sam_S=5/2$ and density $n_S = 0.05/\Omega_0$, where $\Omega_0 = 0.45\times 10^{-22} \text{ cm}^3$ is the unit cell volume in GaAs. a) The carrier and impurity spins closely follow the approximation $P_s(T)=P_s(0)\sqrt{1-(T/T_c)^2}$ from Ref.~\onlinecite{arrott:07C715}. The carrier polarization does not saturate, i.e. $P_{\sigma}(0)\neq 1$, due to a significantly smaller magnetic susceptibility than for the impurities. b) As the effective fields are self-generated, their behavior parallels the $\sqrt{1-(T/T_c)^2}$ shape of the polarizations. The field acting on magnetic impurities is largely due to the second order RKKY-like interaction mediated by carriers, and scales with $T_c$ for varying interaction strengths.}
\label{fig:PolFields}
\end{figure}

We can self-consistently solve for the polarizations of, and average fields acting on magnetic impurities and carriers below $T_c$. The polarization of the magnetic impurities is given by the Brillouin function $\Br_J(x)$:
\begin{equation}
	\label{polS}
	P_S = \Br_{\Sam_S}\left[\frac{\Sam_S \left|\langle B_{S,z} \rangle\right|}{k_BT}\right].
\end{equation}
To a good approximation, the field acting on carriers is equivalent to the first order interaction with magnetic impurities:
\begin{equation}
	\label{Bsig1}
	\left|\langle B_{\sigma,z} \rangle\right| \cong J\Sam_S n_S P_S.
\end{equation}
The field acting on carriers also defines the spin splitting energy $\ve_{F\ua}-\ve_{F\da}$ about the Fermi energy $\ve_F$:
\begin{equation}
	\label{Bsig2}
	\left|\langle B_{\sigma,z} \rangle\right| = \ve_F\left[(1+P_{\sigma})^{2/3} - (1-P_{\sigma})^{2/3}\right].
\end{equation}
Finally, we ensure self-consistency in the generated fields through expectation of a Goldstone mode on the ferromagnetic side of the transition. Indeed, in this broken symmetry state the polarized spins can all rotate together without any energy cost. This mode is identified by setting $q=\omega=0$ in the effective transverse susceptibility and stipulating a divergence:
\begin{multline}
	\label{goldstone}
	1-(J/2)^2\chi^{(0)}_{S_+S_-}(T) \lim_{q\rightarrow 0} \chi^{(0)}_{\sigma_+\sigma_-}(q,0) =\\
	1-(J/2)^2\frac{2\langle S_z \rangle}{\langle B_{S,z} \rangle}\frac{2\langle \sigma_z \rangle}{\langle B_{\sigma,z} \rangle} = 0.
\end{multline}
The resulting polarizations and fields for several values of $J$ are plotted in Fig.~\ref{fig:PolFields}.

It is instructive to examine the strength of the singularity in the effective interactions when $T_c$ is approached. Approximating the non-interacting carrier susceptibility in the paramagnetic phase by its static, small-$q$ form $\chi_{\sigma_z\sigma_z}(q,\omega) \cong -\Sam_{\sigma}^2 N(0) \left(1-q^2/12k_F^2\right)$, we can write the denominator of the scattering amplitude in Eq.~\eqref{Wpar} as
\begin{equation}
	\label{Vqdep}
	1 - J^2 \chi^{(0)}_{S_zS_z}(T)\chi^{(0)}_{\sigma_z\sigma_z}(q,0) \propto \frac{T-T_c}{T_c}+\frac{q^2}{12k_F^2}.
\end{equation}
Then as $T \rightarrow T_c$, we have the same $q^{-2}$ divergence in the interaction strength as a bare Coulomb interaction.

\section{Carrier spin relaxation}
\label{SR}
Typically, spin relaxation is understood to be the evolution of out-of-equilibrium polarized spins into an unpolarized equilibrium state. In a ferromagnetic system where there is an equilibrium spin polarization, spin relaxation refers to the evolution of non-equilibrium spins into the polarized equilibrium state. Thus, rather than picturing the process as decay of polarization, we understand it as restoration of equilibrium polarization.

We focus on two mechanisms of carrier spin relaxation: transfer of spin to magnetic impurities (MI) and Dyakonov-Perel (DP). Under a wide variety of temperatures, scattering sources, and degeneracy, the DP mechanism dominates in III-V semiconductors.\cite{PhysRevB.79.125206,Wu201061} Thus, we find it relevant to include derivations for this spin relaxation mechanism here. As will become evident, however, the inverse relationship between scattering and spin relaxation in the DP mechanism hinders its effectiveness when the interaction that is responsible for carrier-carrier scattering becomes strong and long-ranged, e.g. near the ferromagnetic transition. This leads to a change in dominant carrier spin relaxation mechanism near the ferromagnetic transition. Both the MI and the Elliott-Yafet (EY) mechanisms rely on spin flips in scattering; thus, they scale proportionally to the scattering rate and become more important when the carrier-carrier interaction becomes strong. This point is in agreement with Morandi, \textit{et al.}\cite{PhysRevB.81.155309}, who find that spin-flip-based spin relaxation mechanisms (including a small Elliott-Yafet contribution) are important when spin-dependent scattering is dominant. However, the timescales of these two mechanisms are very different. Since EY depends on the weak spin-orbit field to allow spin flips in interaction, it is generally negligible when MI processes are occurring. This leads us to identify MI as the dominant spin relaxation mechanism near the ferromagnetic transition, and so we neglect the EY mechanism in the following derivations. Nevertheless, even when MI dominates, there is a role left for EY to further relax the total magnetization of impurities plus carriers after the total spin angular momentum has reached equilibrium. This will be qualitatively discussed in Section V. Notice that the enhanced carrier-carrier collision rate near the transition is not in opposition with, but rather the consequence of the critical slowing down of the collective spin dynamics of the carrier-impurity system.

\subsection{Spin flips in scattering}
\label{MISR}
We calculate the rate of spin relaxation due to spin flips in scattering using Keldysh formalism.\cite{neqmbtqs_Stefanucci_vanLeeuwen,haug_qkt} The time rate of change of the average spin density $\langle \sigma_{\kv} \rangle = n_{\sigma} \Sam_{\sigma} \Tr[\rhoh_{\kv}\sigh_z]$ is
\begin{equation}
	\frac{\partial \langle \sigma_{\kv}\rangle}{\partial t} = n_{\sigma} \Sam_{\sigma} \Tr\left[\frac{\partial \rhoh_{\kv}}{\partial t} \sigh_z\right],
\end{equation}
where $\rhoh_{\kv}$ is a $2\times 2$ density matrix in spin space for carriers, $\sigh_z$ is the $z$-Pauli matrix, and $\zh$ has been chosen for the direction of spin polarization. Taking advantage of the relaxation time approximation for $\partial\langle\sigma_{\kv}\rangle/\partial t$ and averaging over all wavevectors $\kv$, the spin relaxation time $\tau_s$ is calculated from
\begin{equation}
	\label{srtscatt_srt}
	\frac{1}{\tau_s} = -\frac{\sum_{\kv}\Tr\left[\partial_t \rhoh_{\kv} \sigh_z\right]}{\sum_{\kv}\Tr\left[\rhoh^1_{\kv} \sigh_z\right]},
\end{equation}
where $\rhoh^1_{\kv}$ is the first order non-equilibrium correction in the linearized density matrix. In the absence of spin-orbit and other external fields, the density matrix evolves according to
\begin{equation}
	\frac{\partial \rhoh_{\kv}}{\partial t} = \Ih_{\kv},
\end{equation}
where $\Ih_{\kv}$ is the $2\times 2$ collision integral in spin space for carriers. The collision integral is expressed in terms of carrier self-energies $\Sigh_{\kv}(\omega)$ and Green's functions $\Gh_{\kv}(\omega)$ as follows:\cite{haug_qkt,PhysRevB.34.6933} 
\begin{multline}
	\label{Collision_Integral}
	\Ih_{\kv} = \frac{1}{2} \int \frac{d\omega}{2\pi} \Big[ \left\{\Sigh_{\kv}^<(\omega),\Gh_{\kv}^>(\omega)\right\}\\
	- \left\{\Sigh_{\kv}^>(\omega),\Gh_{\kv}^<(\omega)\right\} \Big],
\end{multline}
where $\{\cdot\,,\cdot\}$ represents an anti-commutation. In the absence of spin-orbit interactions, small deviations from the equilibrium spin polarization only show up on the diagonal elements of the density matrix. Then, we know \textit{a priori} that $\Gh_{\kv}(\omega)$ and $\Sigh_{\kv}(\omega)$ are diagonal and can write the collision integral per spin-$\alpha$ as
\begin{equation}
	\label{diag_coll_int}
	I_{\kv \alpha} = \int \frac{d\omega}{2\pi} \left[ G^>_{\kv \alpha}(\omega) \Sigma^<_{\kv \alpha}(\omega) - \Sigma^>_{\kv \alpha}(\omega) G^<_{\kv \alpha}(\omega) \right].
\end{equation}

\begin{figure}
\begin{center}
	\includegraphics[width=5cm]{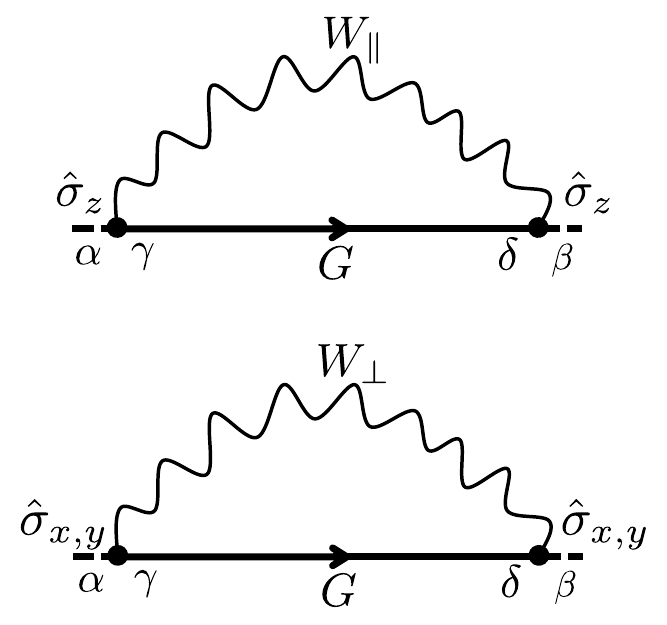}
\end{center}
\caption[Spin indexed GW self-energy]{The GW self-energy includes separate contributions from longitudinal $W_{\parallel}$ and transverse $W_{\perp}$ interactions. Here, $\alpha$, $\beta$, $\gamma$, and $\delta$ are spin states on which the interactions $\sigh_i W \sigh_i$ operate.}
\label{fig:gw_longtrans}
\end{figure}

We evaluate the self-energy in the GW approximation,\cite{0034-4885-61-3-002,bookqtel}
\begin{equation}
	\label{gw_se}
	\Sigma_{\kv \alpha}^{\lessgtr}(\omega) = i\hbar \sum_{\qv} \int \frac{d\Omega}{2\pi} G_{\kv-\qv \alpha}^{\lessgtr}(\omega-\Omega) W_{\qv}^{\lessgtr}(\Omega).
\end{equation}
In allowing for arbitrary spin polarization, we separate the $GW$ spin-space operator into longitudinal and transverse components according to Fig.~\ref{fig:gw_longtrans}: $\widehat{GW} = W_{\parallel} \sigh_z\Gh\sigh_z + W_{\perp} (\sigh_x\Gh\sigh_x + \sigh_y\Gh\sigh_y)$. This allows us to rewrite Eq.~\eqref{gw_se} as
\begin{multline}
	\label{gw_se_lt}
	\Sigma_{\kv \alpha}^{\lessgtr}(\omega) = i\hbar \sum_{\qv} \int \frac{d\Omega}{2\pi} \left[ W_{\parallel\qv}^{\lessgtr}(\Omega)G_{\parallel\kv-\qv \alpha}^{\lessgtr}(\omega-\Omega) \right. \\
	\left.+ 2W_{\perp\qv}^{\lessgtr}(\Omega)G_{\perp\kv-\qv \alpha}^{\lessgtr}(\omega-\Omega) \right],
\end{multline}
where the newly defined longitudinal and transverse Green's functions are $\Gh_{\parallel\kv} = \sigh_z\Gh_{\kv}\sigh_z$ and $\Gh_{\perp\kv} = (1/2)[\sigh_x\Gh_{\kv}\sigh_x + \sigh_y\Gh_{\kv}\sigh_y]$.

The GKB ansatz\cite{haug_qkt} -- an exact relation in non-interacting systems -- is utilized to relate the lesser and greater Green's functions to the density matrices $\rhoh^<_{\kv}=\rhoh_{\kv}$ and $\rhoh^>_{\kv}=\one-\rhoh_{\kv}$, where $\one$ is the $2\times2$ identity matrix:
\begin{equation}
	\label{gkb_rho}
	G_{\kv \alpha}^{\lessgtr}(\omega) = \mp G_{\kv \alpha}^r(\omega)\rho^{\lessgtr}_{\kv \alpha} \pm \rho^{\lessgtr}_{\kv \alpha} G^a_{\kv \alpha}(\omega).
\end{equation}
Notice that we have neglected retardation effects in the density matrix by making it frequency independent. This is commonly referred to as the Markovian approximation and is justified when the time evolution of the distribution function is slow on the time scale of microscopic collisions. We let all non-equilibrium properties reside in $\rhoh_{\kv}$, so $\Gh^{r,a}_k(\omega)$ are taken to be equilibrium Green's functions. Furthermore, given that our aim is to derive a Boltzmann-like collision integral, we retain only the quasiparticle singularities in $\Gh^{r,a}_k(\omega)$, as those are the parts which conserve energy in collisions.\cite{PhysRevLett.73.3439} Under these circumstances, the lesser and greater Green's functions become
\begin{equation}
	\label{Glg_mrt}
	G^{\lessgtr}_{\kv \alpha}(\omega) = \pm i 2\pi \delta(\hbar\omega - \ve_{k\alpha}) \rho_{\kv\alpha}^{\lessgtr}.
\end{equation}
Similarly, we can use the fluctuation dissipation theorem to relate the lesser and greater interactions $W_{\qv}^{\lessgtr}(\omega)$ to the imaginary part of the retarded interaction $\Im m W_{\qv}(\omega)$:\cite{neqmbtqs_Stefanucci_vanLeeuwen}
\begin{equation}
	\label{W_lessgtr}
	W^{\lessgtr}_{\qv}(\omega) = \pm i 2 n_B(\pm\omega) \Im m W_{\qv}(\omega),
\end{equation}
where $n_B(\omega) = (e^{\beta\hbar\omega}-1)^{-1}$ is the Bose-Einstein distribution and $\beta=(k_BT)^{-1}$.

The result of inserting Eqs.~\eqref{gw_se_lt}, \eqref{Glg_mrt}, and \eqref{W_lessgtr} into the collision integral, Eq.~\eqref{diag_coll_int}, is
\begin{align}
	\label{srtscatt_diag_coll_int}
	& I_{\kv \alpha} = 2 \sum_{\qv} \int d\omega\, \Big\{ \Im m W_{\parallel \qv}(\omega) \delta(\ve_{k \alpha} - \ve_{\kv-\qv \alpha} - \hbar\omega) \nonumber \\
	&\quad \left[ n_B(\omega) (1-\rho_{\kv \alpha})\rho_{\kv-\qv \alpha} + n_B(-\omega) \rho_{\kv \alpha} (1-\rho_{\kv-\qv \alpha}) \right] \nonumber \\
	&+ 2\Im m W_{\perp \qv}(\omega) \delta(\ve_{k \alpha} - \ve_{\kv-\qv \bar{\alpha}} - \hbar\omega) \nonumber\\
	&\quad \left[ n_B(\omega) (1-\rho_{\kv \alpha})\rho_{\kv-\qv \bar{\alpha}} + n_B(-\omega) \rho_{\kv \alpha} (1-\rho_{\kv-\qv \bar{\alpha}}) \right] \Big\}.
\end{align}
For a slightly out of equilibrium spin polarization, the distribution $\rho_{\kv\alpha}$ can be linearized as follows:
\begin{equation}
	\label{srtscatt_lin_f}
	\rho_{\kv\alpha} = f^0_{k\alpha} + \alpha \beta\ve_{\alpha} f^0_{k\alpha}(1-f^0_{k\alpha}),
\end{equation}
where $\ve_{\alpha}$ is a small spin perturbation energy, $\alpha=+1$ for spin-up, and $\alpha=-1$ for spin-down. Notice that the shifts $\ve_{\alpha}$ and $\ve_{\bar{\alpha}}$ are related by the condition $\ve_{\alpha} f^0_{k\alpha}(1-f^0_{k\alpha}) = \ve_{\bar{\alpha}} f^0_{k\bar{\alpha}}(1-f^0_{k\bar{\alpha}})$, or equivalently after summing over $\kv$-space, $\ve_{\alpha} N_{\alpha}(0) = \ve_{\bar{\alpha}} N_{\bar{\alpha}}(0)$, meaning that the departure from equilibrium involves only a change in spin density, not a change in carrier density. From a physical standpoint, it is clear that longitudinal interactions will preserve spin polarization, and so we should see that part of the collision integral vanish upon insertion of Eq.~\eqref{srtscatt_lin_f} into Eq.~\eqref{srtscatt_diag_coll_int}. Indeed, this is the case and we only need to evaluate
\begin{multline}
	\label{srtscatt_lin_coll_int}
	I_{\kv \alpha} = -4\alpha\beta\left(\ve_{\alpha}+\ve_{\bar{\alpha}}\right) \sum_{\qv} \int d\omega\, \Im m W_{\perp \qv}(\omega) \\
	\times \delta(\ve_{k \alpha} - \ve_{\kv-\qv \bar{\alpha}} - \hbar\omega) \frac{f^0_{k \alpha}-f^0_{\kv-\qv \bar{\alpha}}}{4\sinh^2(\beta\hbar\omega/2)},
\end{multline}
which is entirely controlled by the transverse part of the interaction.

\begin{figure}
\begin{center}
	\includegraphics[width=0.48\textwidth]{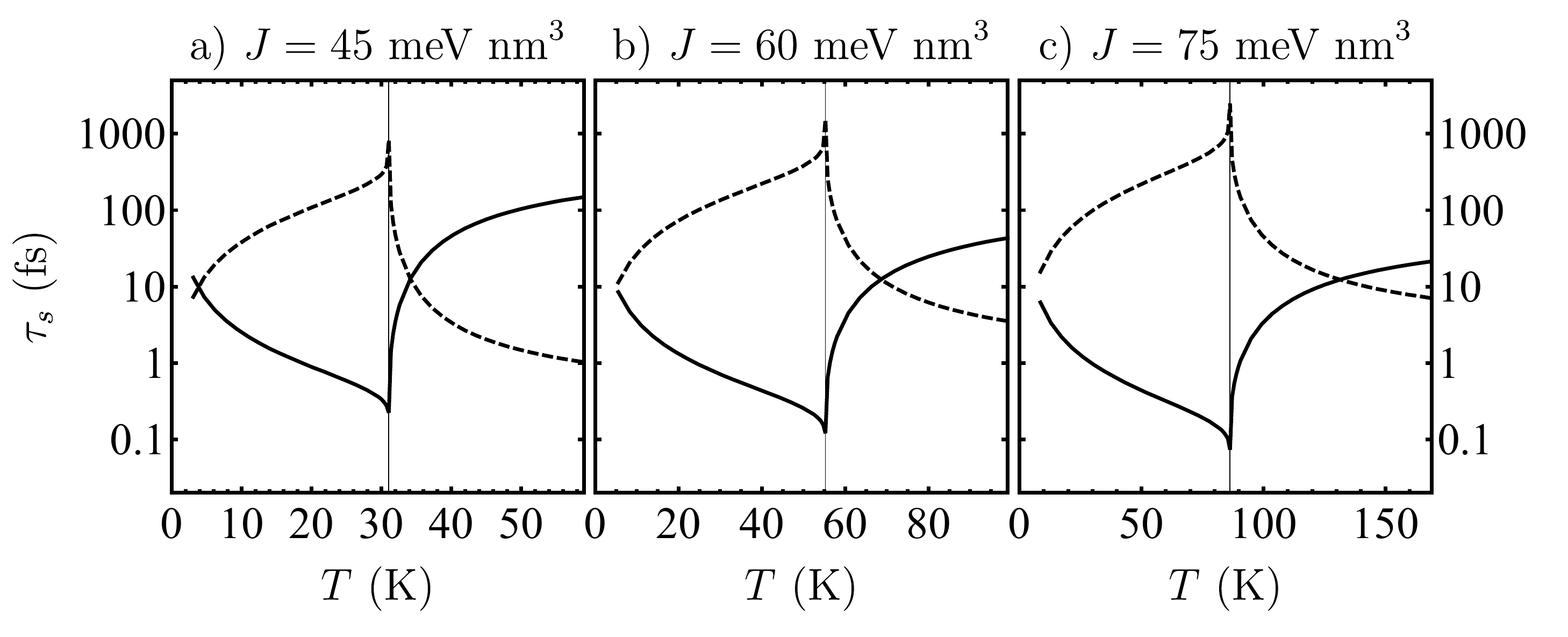}
\end{center}
\caption[Spin relaxation times from spin-flips and Dyakonov-Perel]{The spin-flip based spin relaxation time (solid lines) resulting from the spin dependent interaction is plotted vs. temperature for several values of $J$. For comparison, the Dyakonov-Perel spin relaxation time is shown by the dashed lines. We used GaAs parameters for holes in calculations: $m^*=0.5m_e$, $\alpha_{s.o.}=0.34$, and $E_g = 1.5$ eV. The vertical lines mark the critical temperature in each plot. The short spin relaxation time near the ferromagnetic transition correlates with short spin lifetimes in the spin-flip mechanisms and long spin lifetimes in the Dyakonov-Perel mechanism. Away from the ferromagnetic transition, we see Dyakonov-Perel spin relaxation becoming dominant as is often found in III-V semiconductors.}
\label{fig:SRTComp}
\end{figure}

Returning to the spin relaxation rate defined in Eq.~\eqref{srtscatt_srt}, we can insert expressions for $\rhoh_{\kv}$ and $\Ih_{\kv}$. The rate of spin relaxation from transverse spin scattering events is
\begin{multline}
	\label{srtscatt_final_srt}
	\frac{1}{\tau_s} = \frac{2\beta}{\pi}\left(\frac{1}{N_{\ua}(0)}+\frac{1}{N_{\da}(0)}\right) \\
	\times \sum_{\qv,\alpha} \int d\omega\, \Im m W_{\perp \qv}(\omega) \frac{\Im m\chi_{\bar{\alpha}\alpha}[1,q,\omega]}{4\sinh^2(\beta\hbar\omega/2)},
\end{multline}
where we have introduced the Lindhard-like imaginary response function
\begin{multline}
	\label{chi_ll_function}
	\Im m\chi_{\alpha\beta}[\xi, q, \omega] \equiv \\
	\pi \sum_{\kv} \delta(\ve_{\kv+\qv \beta} - \ve_{k \alpha} - \hbar\omega) \left[f^0_{\kv+\qv \beta}-f^0_{k \alpha}\right]\xi.
\end{multline}
The variable $\xi$ is $1$ in Eq.~(\ref{srtscatt_final_srt}) but will take on more interesting forms in subsequent sections. The spin relaxation time calculated from Eq.~(\ref{srtscatt_final_srt}) is plotted as a dashed line in Fig.~\ref{fig:SRTComp} for three different values of the exchange coupling $J$, above and below the ferromagnetic transition temperature. The sharp drop of $\tau_s$ in the immediate vicinity of the ferromagnetic transition reflects the enhanced spin-flip rate from critical fluctuation in this region.

\subsection{Dyakonov-Perel mechanism}
\label{DPSR}

In the DP mechanism, spins relax due to precession around axes defined by a magnetic field which varies in $\kv$-space. In III-V semiconductors, this field is typically Dresselhaus and/or Rashba spin-orbit. The effect of scattering on this process is usually to inhibit spin relaxation as spins are unable to make full precessions before being relocated in momentum space. This sets a timescale, where the rate of scattering must be faster than the rate of spin precession for spin relaxation to be inhibited, otherwise spins enter a regime of free precessions and the rate of relaxation is directly proportional to the rate of precession. In the limit of fast scattering, spins are unable to make appreciable precessions around the spin-orbit field's axes and spin relaxation is entirely suppressed. This is the scenario we encounter.

The derivation of DP spin relaxation is reviewed in Refs.~\onlinecite{PIKUS198473,ActaPhysSlov.57.565,PhysRevB.83.155205}. When the Hamiltonian includes a contribution from a spin-orbit field $\Omegv_{\kv}$ of the form
\begin{equation}
	\label{H1SO}
	\Hm_{\kv}^1 = \frac{\hbar}{2} \Omegv_{\kv} \cdot \hat{\sigv},
\end{equation}
where $\hat{\sigv}$ is the Pauli matrix vector, the rate of spin relaxation $1/\tau_s$ for carriers due to the DP mechanism is\cite{PhysRevB.83.155205}
\begin{equation}
	\label{tau_s}
	\frac{1}{\tau_s} = \frac{2}{3}\frac{\sum_{\kv} \tau_k^* \langle \Omega_{\kv}^2 \rangle \Tr\left[\rhoh^0_k\sigh_z\right]}{\sum_{\kv} \Tr\left[\rhoh^0_k\sigh_z\right]},
\end{equation}
where $\tau_k^*$ is an effective scattering time that weights collisions according to their ability to randomize the spin precession axis. The angle brackets $\langle \cdot \rangle$ in Eq.~\eqref{tau_s} represent an average over all directions of $\kv$, and in a bulk III-V semiconductor this quantity evaluates to
\begin{equation}
	\label{ang_avg_so_dres}
	\langle{\Omega_{\kv}^2}\rangle = \frac{16}{35} \frac{\alpha_{s.o.}^2 \ve_k^3}{\hbar^2 E_g},
\end{equation}
where $\alpha_{s.o.}$ is the spin-orbit coupling constant and $E_g$ is the band gap energy.

We emphasize here that spin-flip processes are not considered in Eq.~\eqref{tau_s}. This spin relaxation rate is entirely due to spin-orbit coupling. The challenge in evaluating Eq.~\eqref{tau_s} comes from the effective scattering rate $1/\tau_k^*$. An oft-made approximation is to substitute the rate of momentum relaxation in its place. The validity of this approximation in the context of spin-scattering will be discussed in section \ref{DISCUSSION}. In this section we present our calculation of $1/\tau_k^*$.

Written in terms of density ($f^n$) and spin ($f^s$) distributions, the quasi-equilibrium density matrix $\rhoh^0_k$ is
\begin{equation}
	\label{rho0}
	\rhoh^0_k = f^{0n}_k \one + (f^{0s}_k + f^{1s}_k) \sigh_z,
\end{equation}
where
\begin{align*}
	f^{0n}_k &= (f^0_{k\ua}+f^0_{k\da})/2, \\
	f^{0s}_k &= (f^0_{k\ua}-f^0_{k\da})/2, \\
	f^{1s}_k &= (f^1_{k\ua}-f^1_{k\da})/2,
\end{align*}
and $f^1_{k\alpha}$ was defined in Eq.~\eqref{srtscatt_lin_f}. In writing the rate of spin relaxation in Eq.~\eqref{tau_s}, there is an implicit assumption that a small spin polarization has been prepared which will relax. This is the reason for the non-equilibrium $f^{1s}_k$ contribution in Eq.~\eqref{rho0}; it is stable on the short time scale of scattering, but variable on the longer time scale of spin relaxation. Then, while it is necessary to include this term in Eq.~\eqref{tau_s}, it can be dropped when evaluating the collision integral, where we can instead use $\rho^0_{k\alpha} = f^0_{k\alpha}$.

The non-equilibrium part of the density matrix $\rhoh_{1\kv}$ arises from interaction with the spin-orbit field:\cite{PhysRevB.83.155205}
\begin{align}
	\label{rho1} 
	\rhoh^1_{\kv} &= \frac{\tau^*_k}{2i} \left[\Omegv_{\kv}\cdot\hat{\sigv}, \rhoh^0_k\right] \nonumber \\
	&= \tau^*_k f^{1s}_k (\Omegv_{\kv}\times\zh) \cdot \hat{\sigv},
\end{align}
where $[\cdot\,,\cdot]$ represents a commutation and $\hat{\sigv} = \{\sigh_x,\sigh_y,\sigh_z\}$ is the Pauli spin vector. We use a relaxation time approximation to relate $\rhoh^1_{\kv}$ and $\tau^*_k$ to the collision integral $\Ih_{\kv}$ as follows:
\begin{equation}
	\label{rho1_rta}
	-\frac{\rhoh^1_{\kv}}{\tau^*_k} = \Ih_{\kv}.
\end{equation}
One should exercise caution when evaluating this collision integral, though. While $\Ih_{\kv}$ in Eq.~\eqref{Collision_Integral} is still generally applicable, the GKB ansatz for $G_{\kv\alpha\beta}^{\lessgtr}(t_1,t_2)$ is not simply dependent on the difference $t_1-t_2$ when $\alpha \neq \beta$, and so does not lend itself easily to Fourier transform. Instead, we have (see Appendix \ref{GKB_FM}):
\begin{equation}
	\label{gkb_spinindexed}
	G^{\lessgtr}_{\kv\alpha\beta}(t_1,t_2) = \pm i e^{-i\ve_{k\alpha}t_1+i\ve_{k\beta}t_2} \rho^{\lessgtr}_{\kv\alpha\beta}.
\end{equation}
Thus, we opt to start in the time-domain with collision integral\cite{haug_qkt}
\begin{multline}
	\label{t_coll_int}
	I_{\kv}(t) = \int_{-\infty}^t dt'\, \left[ \Sigma^<_{\kv}(t,t')G^>_{\kv}(t',t) + G^>_{\kv}(t,t')\Sigma^<_{\kv}(t',t)\right. \\
	\left.- \Sigma^>_{\kv}(t,t')G^<_{\kv}(t',t) - G^<_{\kv}(t,t')\Sigma^>_{\kv}(t',t) \right].
\end{multline}
The self-energy in the GW approximation is
\begin{multline}
	\label{t_se}
	\Sigma^{\lessgtr}_{\kv}(t_1,t_2) = i \sum_{\qv} \left[ W^{\lessgtr}_{\parallel\qv}(t_1,t_2) G^{\lessgtr}_{\parallel\kv-\qv}(t_1,t_2) \right.\\
	\left. + 2W^{\lessgtr}_{\perp\qv}(t_1,t_2) G^{\lessgtr}_{\perp\kv-\qv}(t_1,t_2) \right].
\end{multline}
After inserting Eqs.~\eqref{gkb_spinindexed} and \eqref{t_se} into Eq.~\eqref{t_coll_int}, the time of observation can be set to $t=0$ and integrals over $t'$ performed. The fluctuation-dissipation theorem from Eq.~\eqref{W_lessgtr} is again utilized for $W^{\lessgtr}(\omega)$ and the collision integral reduces to a strictly off-diagonal matrix:
\begin{align}
	\label{I_w_linearized}
	& I_{\kv\alpha\bar{\alpha}} = -\sum_{\qv,\gamma} \int_{-\infty}^{\infty} d\omega\, \Big\{ \Im m W_{\parallel\qv}(\omega) \delta(\ve_{k\gamma}-\ve_{\kv-\qv\gamma}-\hbar\omega) \nonumber \\
	& \quad \left(\left[n_B(-\omega) + f^0_{\kv-\qv\gamma}\right] \rho^1_{\kv\alpha\bar{\alpha}} - \left[n_B(\omega) + f^0_{k\gamma}\right] \rho^1_{\kv-\qv\alpha\bar{\alpha}}\right) \nonumber \\
	&- 2\Im m W_{\perp\qv}(\omega) \delta(\ve_{k\bar{\gamma}}-\ve_{\kv-\qv\gamma}-\hbar\omega) \nonumber \\
	& \quad \left[n_B(-\omega) + f^0_{\kv-\qv\gamma}\right] \rho^1_{\kv\alpha\bar{\alpha}} \Big\}.
\end{align}
This is an appropriate time to re-introduce the relaxation time approximation from Eq.~\eqref{rho1_rta} and substitute the non-equilibrium density matrix elements from Eq.~\eqref{rho1}. Taking advantage of $\ve_{\alpha} f^0_{k\alpha}(1-f^0_{k\alpha}) = \ve_{\bar{\alpha}} f^0_{k\bar{\alpha}}(1-f^0_{k\bar{\alpha}})$, the kinetic equation simplifies to
\begin{align}
	\label{I_w_sinh}
	& \frac{1}{2}\sum_{\gamma} \ve_{\gamma} f^0_{k\gamma}(1-f^0_{k\gamma}) \Omega_{\kv+} = \nonumber \\
	& \sum_{\qv,\gamma} \int_{-\infty}^{\infty} d\omega\, \bigg\{ \Im m W_{\parallel\qv}(\omega) \delta(\ve_{k\gamma}-\ve_{\kv-\qv\gamma}-\hbar\omega) \nonumber \\
	& \quad \beta\ve_{\gamma} \frac{f^0_{k\gamma}-f^0_{\kv-\qv\gamma}}{4\sinh^2(\beta\hbar\omega/2)} \left(\tau^*_k\Omega_{\kv+} + \tau^*_{\kv-\qv}\Omega_{\kv-\qv+}\right) \nonumber \\
	&+ 2 \Im m W_{\perp\qv}(\omega) \delta(\ve_{k\bar{\gamma}}-\ve_{\kv-\qv\gamma}-\hbar\omega) \nonumber \\
	& \quad \beta\ve_{\bar{\gamma}} \frac{f^0_{k\bar{\gamma}}-f^0_{\kv-\qv\gamma}}{4\sinh^2(\beta\hbar\omega/2)}\tau^*_k\Omega_{\kv+} \bigg\},
\end{align}
where $\Omega_{\kv+} = \Omega_{\kv,y} + i \Omega_{\kv,x}$.

This integral equation can be solved exactly using the methods of Sykes and Brooker,\cite{Sykes19701} but given the slowly varying nature of $\tau^*_k$ around the Fermi level in the degenerate regime, it is a reasonable approximation to treat it as a constant evaluated at $\ve_F$ and extract it from the collision integral. Then, we only need to specify the spin-orbit field and solve for the effective scattering rate.

\begin{figure}
\begin{center}
	\includegraphics[width=0.48\textwidth]{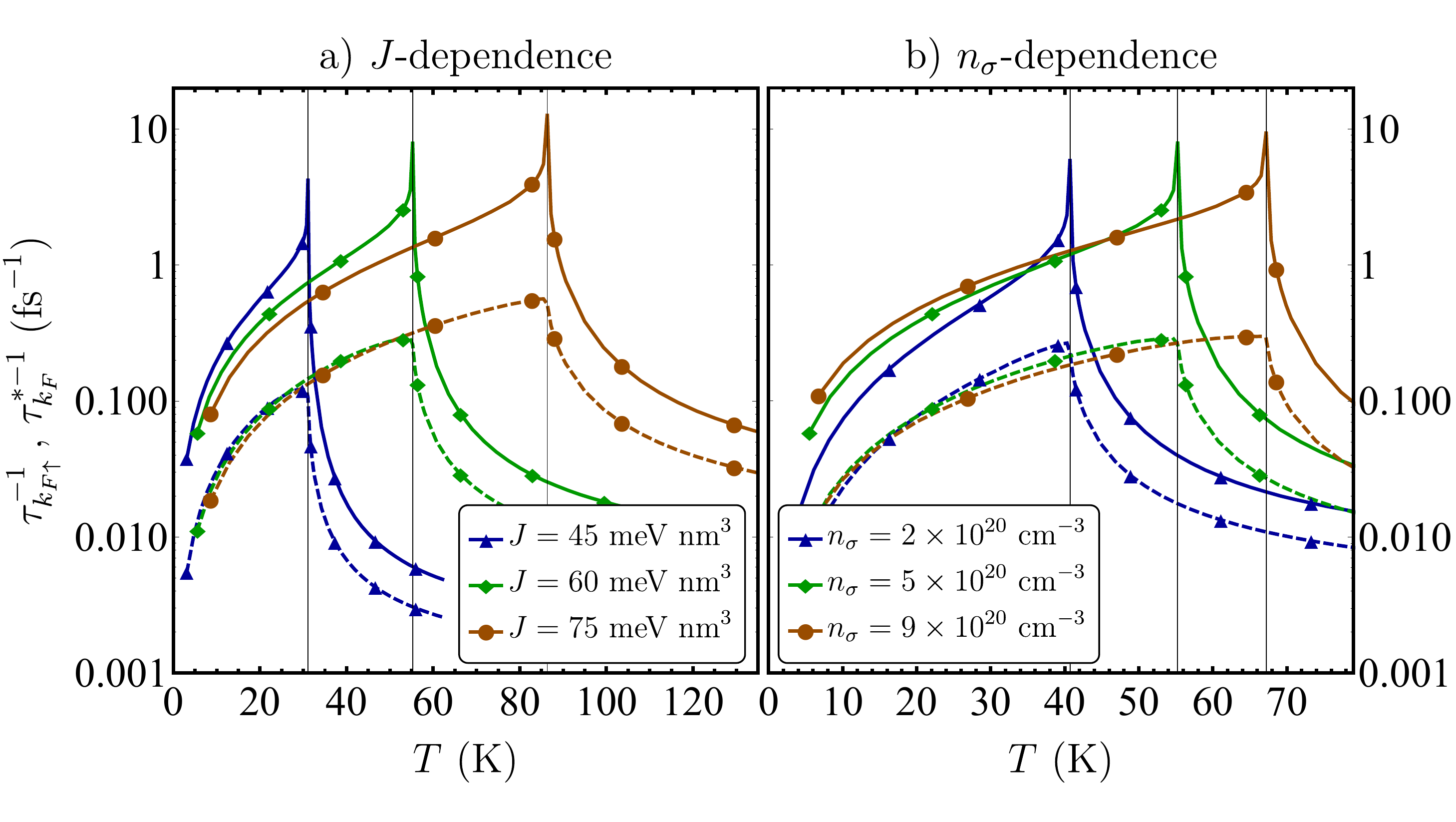}
\end{center}
\caption[Effective scattering rate compared to momentum relaxation]{(Color online) The effective scattering rate used in Dyakonov-Perel spin relaxation calculations (solid lines) is compared to the momentum relaxation rate (dashed lines) for varying a) exchange interaction strengths and b) carrier densities. The solid vertical lines mark critical temperatures. The effective scattering rates diverge at the ferromagnetic transition, whereas the rate of momentum relaxation does not. This ultimately comes from zero-momentum transfer processes in spin-spin scattering events that affect Dyakonov-Perel spin relaxation, but not momentum relaxation.}
\label{fig:ESRvMRT}
\end{figure}

The Dresselhaus field\cite{PhysRev.100.580} components, written in terms of spherical harmonics $Y_{l,m}(\vt,\vp)$ are
\begin{subequations}
	\label{dresselhaus}
\begin{align}
	\label{omega_x}
	\Omega_{\kv,x} &= \Omega_0 k^3 \left[ \sqrt{\pi/21} \left(Y_{3,1}(\vt,\vp)-Y_{3,-1}(\vt,\vp)\right) \right. \nonumber \\
	&\quad \quad \quad \left. + \sqrt{\pi/35} \left(Y_{3,3}(\vt,\vp)-Y_{3,-3}(\vt,\vp)\right)\right], \\
	\label{omega_y}
	\Omega_{\kv,y} &= \Omega_0 k^3 \left[ i\sqrt{\pi/21} \left(Y_{3,1}(\vt,\vp)+Y_{3,-1}(\vt,\vp)\right) \right. \nonumber \\
	&\quad \quad \,\,\,\,\, \left. - i \sqrt{\pi/35} \left(Y_{3,3}(\vt,\vp)+Y_{3,-3}(\vt,\vp)\right)\right], \\
	\label{omega_z}
	\Omega_{\kv,z} &= \Omega_0 k^3 \sqrt{8\pi/105} \left(Y_{3,2}(\vt,\vp)+Y_{3,-2}(\vt,\vp)\right),
\end{align}
\end{subequations}
where $\Omega_0 = \alpha_c \hbar^2 / \sqrt{2m_c^3E_g}$. These forms allow us to take advantage of the addition theorem for spherical harmonics, resulting in the following useful identity:
\begin{equation}
	\label{VOmega_to_Legendre}
	\int \frac{d\Omega'}{4\pi} V_{\kv-\kv'} \Omega_{\kv',j} = \Omega_{\kv,j} \left(\frac{k'}{k}\right)^3 \int \frac{d\Omega'}{4\pi} V_{\kv-\kv'} P_3(\cos\vartheta'),
\end{equation}
where $d\Omega'=d(\cos\vartheta')d\varphi'$ and $j$ can be any of $x$, $y$, or $z$. Pushing $\sum_{\kv} \Omega_{\kv,+}^*/k^6$ onto both sides of Eq.~\eqref{I_w_sinh}, we can extract the response function $\Im m\chi_{\alpha\beta}[\xi^{(dp)}_q,q,\omega]$ defined in Eq.~\eqref{chi_ll_function}, where 
\begin{equation}
	\xi^{(dp)}_q = \left[\frac{k}{\left|\kv+\qv\right|}\right]^3 P_3\left[\frac{\kv\cdot\left(\kv+\qv\right)}{k\left|\kv+\qv\right|}\right].
\end{equation}
Notice only the magnitude $|\qv|$ is necessary to evaluate $\Im m\chi_{\alpha\beta}[\xi^{(dp)}_q,q,\omega]$ since $\kv$ is eventually integrated over all directions and can be measured relative to some arbitrary direction of $\qv$. The final result for the effective scattering rate due to a spin dependent interaction in the degenerate regime is
\begin{widetext}
\begin{equation}
	\label{tau_star_final}
	\frac{1}{\tau^*_{k_F}} = \frac{\beta}{\pi} \sum_{\qv, \gamma} \int_{-\infty}^{\infty} d\omega\, \bigg[ \frac{\Im m W_{\parallel\qv}(\omega)}{N_{\gamma}(0)} \frac{\Im m\chi_{\gamma\gamma}[1,q,\omega] + \Im m\chi_{\gamma\gamma}[\xi^{(dp)}_q,q,\omega]}{4\sinh^2(\beta\hbar\omega/2)} + \frac{2\Im m W_{\perp\qv}(\omega)}{N_{\bar{\gamma}}(0)} \frac{\Im m\chi_{\bar{\gamma}\gamma}[1,q,\omega]}{4\sinh^2(\beta\hbar\omega/2)} \bigg].
\end{equation}
The modified response function $\Im m\chi_{\alpha\beta}[\xi^{(dp)}_q,q,\omega]$ can be evaluated analytically. It is rather cumbersome to write in full, so we provide the following form before the final integration is performed:
\begin{align}
	\label{imchi3}
	\Im m\chi_{\alpha\beta}[\xi^{(dp)}_q, q, \omega] = &\frac{\pi N_{\beta}(0)}{4\bar{q}_{\beta}} \Theta\left[1-\nu_{\beta+}^2\right] \int_{\nu_{\beta+}^2}^{1} dx\, \left|\frac{x-\eta_{\beta}}{x}\right|^{3/2} P_3\left[\frac{2x-\eta_{\beta}-\bar{q}_{\beta}^2}{2\sqrt{x^2-x\eta_{\beta}}}\right] \nonumber \\
	- &\frac{\pi N_{\alpha}(0)}{4\bar{q}_{\alpha}} \Theta\left[1-\nu_{\alpha-}^2\right] \int_{\nu_{\alpha-}^2}^{1} dx\, \left|\frac{x}{x+\eta_{\alpha}}\right|^{3/2} P_3\left[\frac{2x+\eta_{\alpha}-\bar{q}_{\alpha}^2}{2\sqrt{x^2+x\eta_{\alpha}}}\right],
\end{align}
where
\begin{align*}
	\bar{q}_{\gamma} &= q/k_{F\gamma}, \\
	\nu_{\gamma\pm} &= \frac{\hbar\omega + (\beta-\alpha)\Delta/2}{\hbar qv_{F\gamma}} \pm \frac{q}{2k_{F\gamma}}, \\
	\eta_{\gamma} &= \frac{\hbar\omega+(\beta-\alpha)\Delta/2}{\ve_{F\gamma}},
\end{align*}
and $\Delta$ is the same spin splitting defined in Eq.~\eqref{Bsig2}. Note that although $\alpha$ and $\beta$ are not explicitly indexed in $\nu_{\gamma\pm}$ and $\eta_{\gamma}$, their ordering in $\beta-\alpha$ is based on direct substitution of these quantities into Eq.~\eqref{imchi3}. To be explicit, when evaluating $\Im m \chi_{\beta\alpha}$, the order changes to $\alpha-\beta$. For comparison, the effective scattering rate obtained when a spin-independent interaction $W_{\qv}$ is used is
\begin{equation}
	\label{tau_star_charge}
	\frac{1}{\tau^*_{k_F}} = \frac{\beta}{\pi} \sum_{\qv, \gamma} \int_{-\infty}^{\infty} d\omega\, \frac{\Im m W_{\qv}(\omega)}{N_{\gamma}(0)} \frac{\Im m\chi_{\gamma\gamma}[1,q,\omega] - \Im m\chi_{\gamma\gamma}[\xi^{(dp)}_q,q,\omega]}{4\sinh^2(\beta\hbar\omega/2)} \quad \left(\genfrac{}{}{0pt}{}{\text{spin-indep.}}{\text{interaction}}\right).
\end{equation}

In Fig.~\ref{fig:ESRvMRT} we plot the calculated values of $1/\tau_{k_F}^*$ vs. temperature for different values of exchange coupling and carrier concentration. We also plot the ordinary momentum relaxation rate which is related to the Drude resistivity $\rho_{\alpha\beta}$ through $\tau_{\alpha}^{-1} = \sum_{\beta} n_{\beta}e^2\rho_{\alpha\beta}/m^*$. A standard calculation yields
\begin{equation}
	\label{mrt_final}
	\frac{1}{\tau_{k_{F\alpha}}} = \frac{\beta}{\pi N_{\alpha}(0)} \sum_{\qv} \int_{-\infty}^{\infty} d\omega\, \bigg[ \Im m W_{\parallel \qv}(\omega) \frac{\hbar\omega + \ve_q}{\ve_{F\alpha}} \frac{\Im m \chi_{\alpha\alpha}[1,q,\omega]}{4\sinh^2(\beta\hbar\omega/2)} \nonumber \\ + 2 \Im m W_{\perp \qv}(\omega) \frac{\hbar\omega + \ve_q + \alpha \Delta}{\ve_{F\alpha}} \frac{\Im m \chi_{\bar{\alpha}\alpha}[1,q,\omega]}{4\sinh^2(\beta\hbar\omega/2)} \bigg].
\end{equation}
\end{widetext}

We note that this formula generalizes a previous expression that was derived for the contribution of spin fluctuations to the momentum relaxation rate (see Eq.~(18) of Ref.~\onlinecite{PhysRevB.80.205202}). The essential difference is that our formula takes into account inelasticity of the scattering, whereas Ref.~\onlinecite{PhysRevB.80.205202} treated the spin fluctuations in a quasi-static limit, which misses the critical enhancement of the scattering rate at the ferromagnetic transition. The quasi-static formula is obtained from the present formula by the replacements
\begin{equation}
	\label{QuasiStatic}
	\Im m W_{\parallel (\perp)\qv}(\omega) \to -\beta |U_{\parallel(\perp)}(\qv)|^2\omega\delta(\omega),
\end{equation}
where $|U_{\parallel(\perp)}(\qv)|^2$ is the squared matrix element of the spin-dependent part of the static electron-impurity interaction (longitudinal or transverse) between the initial and the final state of the carrier. Note that one cannot set $\omega \delta(\omega) \simeq 0$ in Eq.~\eqref{QuasiStatic} because that distribution will be multiplied, in Eq.~\eqref{mrt_final}, by quantities that tend to infinity for $\omega \to 0$.

\section{Discussion}
\label{DISCUSSION}
Our expression~\eqref{tau_star_final} for the effective scattering rate in the DP mechanism has some noteworthy features that make it qualitatively different from the previous known result in Eq.~\eqref{tau_star_charge}, which holds for spin-independent interactions. In particular, notice that there is a sign difference in front of $\Im m\chi_{\gamma\gamma}[\xi^{(dp)}_q,q,\omega]$. The significance of this sign is easier to understand if we look back to Eq.~\eqref{I_w_sinh}. In the longitudinal term, after extracting $\tau^*_{k_F}$ as a constant, we have a factor $(\Omega_{\kv+} + \Omega_{\kv-\qv+})$. This should be compared to $(\Omega_{\kv+} - \Omega_{\kv-\qv+})$ which appears when a spin-independent interaction is used. The minus sign in the second case bears some similarity to the $(1-\cos\theta)$ factor that appears in the standard expression for momentum relaxation. A similar factor can be extracted from $(\Omega_{\kv+} - \Omega_{\kv-\qv+})$ and has the form $[1-P_3(\cos\theta)]$, where $P_3(x)$ is the $n=3$ Legendre polynomial.\cite{ActaPhysSlov.57.565,PIKUS198473,PhysRevB.83.155205} For the case of momentum relaxation, when there is no relative angle $\theta$ between incoming and outgoing momenta in a scattering event, there is no contribution to momentum relaxation. The analogy for the effective scattering rate is that zero-momentum transfers do not relocate particles in a $\kv$-dependent magnetic field, so spins can continue to precess around the same magnetic field axis and those scattering events do not reduce the rate of spin relaxation. This analogy is not appropriate for spin-dependent scattering, where processes with zero momentum transfer may still change spin orientation. In fact, what we find is a sort of generalized DP spin relaxation, where spin-dependent and spin-independent scattering should both be taken into account for the overall disruption of spin precessions in a spin-orbit field.

The effective scattering rate from Eq.~\eqref{tau_star_final} is compared to the momentum relaxation rate from Eq.~\eqref{mrt_final} in Fig.~\ref{fig:ESRvMRT} for varying interaction strengths and carrier densities. Unlike previous reports, the effective scattering rate is found to be generally larger than the momentum relaxation rate.\cite{PhysRevB.83.155205} This is consistent with the idea that more scattering events disrupt spin relaxation than affect momentum relaxation. Especially near the critical temperature $T_c$, notice that momentum relaxation has a finite peak whereas the effective scattering rate does not -- there is no vanishing vertex correction in the effective scattering rate to control this divergence. This is an indication that spin exchange and spin flip processes become very important near the critical temperature.

An interesting consequence of the effective interaction used in this paper is that the peak rates of momentum relaxation at $T=T_c$ only indirectly depend on $J$ through the temperature dependence in $1/\sinh^2(\beta\hbar\omega/2)$, and not through the effective interaction $W_{\qv}(\omega)$. This is most easily understood by approaching the ferromagnetic transition from the paramagnetic side. The non-interacting spin response for impurities can be cast as $\chi^{(0)}_{S_zS_z}(T) = -(T_c/T) [J^2\Sam_{\sigma}^2 N(0)]^{-1}$, so that the effective interaction reduces to
\begin{equation}
	\label{Wparamagnetic}
	W_{\qv}(\omega) = -\left[\Sam_{\sigma}^2N(0)(T/T_c)+\chi^{(0)}_{\sigma_z\sigma_z}(q,\omega) \right]^{-1}.
\end{equation}
At $T=T_c$, there is neither an explicit dependence on $J$ nor on $T_c$; we are left with an interaction that only depends on the non-interacting carrier response. The effect of this is $1/\sinh^2(\beta\hbar\omega/2)$ plays the primary role for the temperature dependence. In the small frequency limit, this factor goes as $T^2$. Then, higher critical temperatures have greater scattering rates associated with them; this is verified by Fig.~\ref{fig:ESRvMRT}a. On the other hand, the heights of these scattering peaks are largely independent of the carrier density $n_{\sigma}$, seen in Fig.~\ref{fig:ESRvMRT}b. The primary contribution of carrier density comes from the cut-off frequency $\hbar\omega < \ve_{F\gamma}$ in each of the response functions.

Turning our attention to spin relaxation, the expressions derived for DP spin relaxation rely on the assumption that scattering lifetimes are shorter by roughly an order of magnitude than spin lifetimes. Else, spins are free to precess in a spin-orbit field and scattering does not appreciably affect the rate of spin relaxation.\cite{ActaPhysSlov.57.565,spintronics_dyakonov_book} This is especially relevant for hole spins in III-V semiconductors, as they experience stronger spin-orbit interactions than electrons.\cite{PhysRevLett.89.146601} In Fig.~\ref{fig:SRTComp}, we plot several spin relaxation rates for different exchange interaction strengths $J$. It can be quickly verified by a comparison of timescales that the scattering rates in Fig.~\ref{fig:ESRvMRT} are indeed fast enough to inhibit spin relaxation in the DP mechanism. Near the ferromagnetic transition, the mechanism becomes quite ineffective.

An interesting consequence of carrier spins relaxing by spin exchange with magnetic impurities is the total spin of the system is conserved, which is obviously not the case for spin-orbit based spin relaxation. Near the ferromagnetic transition, this total-spin-conserving relaxation mechanism is very efficient and a quasi-equilibrium spin polarization is reached quickly. From here, spin-orbit based spin relaxation continues to relax carrier spins at a much slower rate until a final stage of equilibrium is reached. This two-stage relaxation process should be observable by a kink in the time evolution of the magnetization. From simple calculations reported in Appendix~\ref{EQ_MI}, we predict this quasi-equilibrium state to have a magnetization in the $\zh$-direction of
\begin{equation}
	M_{z,eq} = \frac{(\chi_{\sigma} g_S+\chi_s g_{\sigma})^2}{(\chi_{\sigma} g_S^2+\chi_s g_{\sigma}^2)(\chi_{\sigma}+\chi_s)} M_{z,i},
\end{equation}
where $M_{z,i}$ is the initial magnetization. The magnetic susceptibilities here do not have $g$-factors absorbed. This highlights the importance of the $g$-factors in defining the quasi-equilibrium state.

Further comparing the spin relaxation mechanisms, a reasonable question to ask is whether the rates of MI, DP, and EY can be summed according to Matthiessen's rule. This would not be appropriate, as the MI mechanism relaxes spins to a different equilibrium state than DP and EY. Especially considering the timescales at which each spin relaxation mechanism operates near the ferromagnetic transition, it makes more sense to order the events as MI spin relaxation to reach a quasi-equilibrium state, and then EY + DP spin relaxation to bring the system to its final equilibrium state where the total spin of the system has decreased.

\section{Conclusion}
\label{CONCLUSION}
We worked with a Zener model of dilute magnetic semiconductors, which exhibits a ferromagnetic transition due to itinerant carriers interacting via spin exchange with localized magnetic impurities. We used this model to derive effective mean-field spin interactions between carriers for many-body calculations. From here, we found analytic expressions for spin-flip spin relaxation, due to exchange of spin between carriers and magnetic impurities, and Dyakonov-Perel. These expressions are valid for degenerate carriers with arbitrary spin polarization and can be used with a variety of spin-dependent dynamic interactions.

We also studied the relative effectiveness of these spin relaxation mechanisms near a ferromagnetic instability. In the Dyakonov-Perel mechanism, spin fluctuations act as effective scatterers, significantly inhibiting the rate of spin relaxation by precession in a spin-orbit field. When this occurs, spin-flip based spin relaxation mechanisms become dominant. When these spin-flips are due to exchange with magnetic impurities, the total spin of the system is conserved and two stages of spin relaxation are expected: fast relaxation occurs due to spin exchange and leads to a quasi-equilibrium polarization; this is followed by slower relaxation which dissipates spin to a spin-orbit field.

\section{Acknowledgments}
\label{ACK}
This work was supported by the National Science Foundation under grant number DMR-1104788. We would like to thank Dr. Ilya Tokatly for extensive and enlightening discussions on spin-dependent interactions in the collision integral for a Fermi liquid.

\appendix
\section{GKB ansatz for a general spin-indexed Green's function}
\label{GKB_FM}
In the time-domain, the GKB ansatz is written\cite{haug_qkt}
\begin{multline}
	\hat{G}^{\lessgtr}_{\kv}(t_1,t_2) = \\
	i \hat{G}^r_{\kv}(t_1,t_2)\hat{G}^{\lessgtr}_{\kv}(t_2,t_2) - i\hat{G}^{\lessgtr}_{\kv}(t_1,t_1)\hat{G}^a_{\kv}(t_1,t_2).
\end{multline}
We consider here Green's functions which are matrices in spin-space. The retarded and advanced Green's functions are equilibrium quantities, diagonal in spin-space:
\begin{equation}
	G^{r/a}_{\kv\alpha\alpha}(t_1,t_2) = \mp i \Theta[\pm(t_1-t_2)] e^{-i\ve_{k\alpha}(t_1-t_2)}.
\end{equation}
When the non-equilibrium time-diagonal Green's functions are assumed to be diagonal in spin-space, they can be considered time-independent under the Markovian approximation. If off-diagonal elements are expected, there remains a time propagator which depends on the energy difference between spin states:
\begin{equation}
	G^{\lessgtr}_{\kv\alpha\beta}(t,t) = \pm i \rho^{\lessgtr}_{\kv\alpha\beta} e^{-i\ve_{k\alpha}t+i\ve_{k\beta}t},
\end{equation}
where $\rho$ is a density matrix, $\rho^<=\rho$, and $\rho^>=1-\rho$. Then, the spin-indexed GKB ansatz cannot be simply written in terms of $t_1-t_2$, but instead has the following form:
\begin{equation}
	\label{gkb_final}
	G^{\lessgtr}_{\kv\alpha\beta}(t_1,t_2) = \pm i e^{-i\ve_{k\alpha}t_1+i\ve_{k\beta}t_2} \rho^{\lessgtr}_{\kv\alpha\beta}.
\end{equation}

\section{Quasi-equilibrium magnetization}
\label{EQ_MI}
We seek the intermediate state of spin relaxation where carriers have quickly drained their spin into magnetic impurities and begin a slower loss of spin to orbital angular momentum via spin-orbit based spin relaxation mechanisms. Spin-exchange interactions between carriers and magnetic impurities preserve the total spin $\langle\sigv\rangle + \langle\Sv\rangle$, but as these particles in general have different $g$-factors, the magnetization is not preserved. In what follows, we consider the $\zh$-components of all quantities without explicitly denoting this.

The equilibrium spin densities are determined by minimizing the free energy
\begin{multline}
	F_B\left(\langle \sigma \rangle, \langle S \rangle\right) = \\
	F\left(\langle \sigma \rangle, \langle S \rangle\right) - \mu_B g_{\sigma}\langle \sigma \rangle B - \mu_B g_S \langle S \rangle B
\end{multline}
with respect to $\langle \sigma \rangle$ and $\langle S \rangle$ at constant field $B$. When we remove the magnetic field, we must minimize the free energy subject to the constraint $\langle \sigma \rangle + \langle S \rangle = constant$. This constant is enforced by a Lagrange multiplier $\wt{B}$ which couples with equal strength to each spin. In other words, we have physical coupling $-\mu_B B (g_{\sigma}\langle \sigma \rangle + g_S\langle S \rangle)$ and ``Lagrange coupling" $-\mu_B \wt{B} (\langle \sigma \rangle + \langle S \rangle)$. The magnetic response of the $\sigma$ or $S$ spins to $\tilde B$ is exactly the same as the magnetic response of these spins to physical fields $B_{\sigma} = \tilde{B}/g_{\sigma}$ and $B_S = \tilde{B}/g_S$.

Our system starts slightly out equilibrium so that the magnetization is initially
\begin{align}
	M_i &= \mu_B g_{\sigma} \langle \sigma \rangle_i + \mu_B g_S \langle S \rangle_i \nonumber \\
	&= -(\chi_{\sigma} + \chi_S) B,
\end{align}
where we have used $\mu_B g_{\sigma} \langle \sigma \rangle_{i} = - \chi_{\sigma} B$ and $\mu_B g_{S} \langle S \rangle_{i} = - \chi_{S} B$. Spin exchanges between carriers and magnetic impurities until equilibrium is reached:
\begin{equation}
	M_{eq} = \mu_B g_{\sigma} \langle \sigma \rangle_{eq} + \mu_B g_S \langle S \rangle_{eq},
\end{equation}
where $\mu_B g_{\sigma} \langle \sigma \rangle_{eq} = - \chi_{\sigma} (\tilde B/g_{\sigma})$ and $\mu_B g_{S} \langle S \rangle_{eq} = - \chi_{S} (\tilde B/g_S)$. The Lagrange multiplier field $\tilde{B}$ is determined by the condition
\begin{equation}
	\langle \sigma \rangle_{eq} +\langle S \rangle_{eq}=\langle \sigma \rangle_{i} +\langle S \rangle_{i},
\end{equation}
yielding
\begin{equation}
	\tilde B = g_Sg_{\sigma}\frac{\chi_{\sigma} g_S+\chi_s g_{\sigma}}{\chi_{\sigma} g_S^2+\chi_s g_{\sigma}^2}B,
\end{equation}
and finally
\begin{align}
	M_{eq} &= -\frac{(\chi_{\sigma} g_S+\chi_s g_{\sigma})^2}{\chi_{\sigma} g_S^2+\chi_s g_{\sigma}^2}\mu_B B \nonumber \\
	&= \frac{(\chi_{\sigma} g_S+\chi_s g_{\sigma})^2}{(\chi_{\sigma} g_S^2+\chi_s g_{\sigma}^2)(\chi_{\sigma}+\chi_s)} M_i.
\end{align}

\end{document}